\documentclass[11pt]{article}
\usepackage{amsmath,amssymb,epsfig,latexsym}
\begin{document}

\begin{center}
{\bf TWISTOR AND "WEAK" GAUGE STRUCTURES IN THE FRAMEWORK OF} 
{\bf QUATERNIONIC ANALYSIS}\vskip 1cm
{Vladimir~V~Kassandrov and Joseph A Rizcallah} \vskip 1 cm
\end{center}
{\footnotesize Department of General Physics, People's Friendship University of 
Russia,Ordjonikidze 3, 117419 Moscow, Russia\\ 
E-mail: vkassan@sci.pfu.edu.ru}\vskip 1cm
\noindent
{\footnotesize {\bf Abstract.} The earlier proposed conditions of {\it (bi)quaternionic 
differentiability} are nonlinear, give rize to the 2-spinor and the self-dual 
gauge structures and may be considered as the {\it generating system of 
equations} (GSE) with respect to the source-free Maxwell, Yang-Mills and 
eikonal equations. We present the  general solution of the 
GSE in terms of twistor variables, analize its rather specific gauge 
symmetry and demonstrate the relation of GSE to the equations of 
shear-free null congruences and, consequently, - to  effective
metrics of Kerr-Shild type. The concept of particles as singularities 
of physical fields associated with the solutions of GSE is developed}\vskip1cm
\noindent
{\footnotesize {\bf KEY WORDS :} Noncommutative analysis; quaternionic 
differentiability; twistors; self-duality; shear-free congruences; 
Kerr-Schild metrics; singularities.}\vskip1cm
\noindent
{\footnotesize PASC numbers: 0240, 0470B, 1130P, 1125M}

\section{Introduction}
\label{sec1}
We present here an effectively nonlinear, non-Lagrangian system of PDE 
which follow from an invariant differential-form condition
\begin{equation}\label{GSE}
d\xi=\Phi dX \xi
\end{equation}
for a 2-spinor $\xi(x)$ and a $Mat(2,\mathbb{C})$-valued gauge $\Phi(x)$ 
fields. Condition (\ref{GSE}) has been proposed in ~\cite{kass1,kass2} in the 
general framework of {\it algebrodynamical} approach to field theory
~\footnote{In algebrodynamics one claims to derive field equations from only the 
intrinsic properties of fundamental mathematical structures, in particular 
from exeptional quaternion-type algebras.}. 
In (\ref{GSE}) $d$ is an operator of external differentiation, and by 
$X$ a hermitian $2\times 2$ matrix of the Minkowski space-time coordinates is denoted. 
System (\ref{GSE}) originates as a particular case of the {\it differentiability 
conditions} for functions of {\it biquaternionic} ($\mathbb{B}$-) variable (we 
recall that the algebra of complex quaternions $\mathbb{B}$ is isomorphic to the 
full algebra of $2\times 2$ complex matrices, $\mathbb{B}\cong Mat(2,\mathbb{C})$). 
Geometrically, it defines a spinor field  covariantly constant with respect 
to an effective $\mathbb{B}$-valued connection 1-form 
$\Omega=\Phi dX$ ~\footnote{In a 4-vector representation this connection 
1-form gives rize to an effective Weyl-Cartan geometry with the  Weyl nonmetricity 
vector and the pseudotrace of the torsion tensor being proportional to each other, 
see section 5.}. 

Remarkably, system (\ref{GSE}) is overdetermined, so that both the 2-spinor 
and the gauge fields are to be found from it in a self-consistent manner. 
The integrability conditions for (\ref{GSE}) take the form 
\begin{equation}\label{comp22}
dd\xi \equiv 0 = (d\Phi - \Phi dX \Phi)\wedge dX \xi \equiv R\xi
\end{equation}
and, as shown below, lead to self-duality of the curvature 2-form $R$ 
of the connection $\Omega$. Consequently, the source-free Maxwell and Yang-Mills 
equations are satisfied identically on the solutions to (\ref{GSE}), for the 
trace and the trace-free parts of the curvature $R$ respectively. For this 
reason, in what follows system (\ref{GSE}) is referred to as the {\it 
generating system of equations} (GSE).

Moreover, the 4-eikonal and the linear wave equation do hold both for each 
spinor component $\{\xi_{A}\},~A=0,1$ (the latter only for the quotient of 
components). Also, a null geodesic congruence which may be constructed via its 
tangent 4-vector $k_\mu =\xi_A \xi_{A^\prime}$ turns out to be {\it shear-free} on the solutions 
of (\ref{GSE}) and, therefore, the {\it Kerr theorem} ~\cite{pr} may be 
applied to the GSE. On this track,  general analytical solution to (\ref{GSE}) may 
be obtained in terms of {\it twistor variables} $T=\{\xi, iX\xi\}$ and in an 
implicit algebraic form 
\begin{equation}\label{GENSOL}
\Pi^{(C)} (\xi, iX\xi)=0,
\end{equation}
where $\{\Pi^{(C)}\},~C=0,1$ are two arbitrary and independent holomorphic 
functions of 4 twistor components. If some $\Pi^{(C)}$ are chosen, a solution 
$\xi_A(x)$ can be extracted from the system of two algebraic equations (\ref{GENSOL}), 
and all the analytical solutions to the GSE can be obtained in this way.

Expression (\ref{GENSOL}) generalizes the Kerr theorem which deals with  
only one homogeneous function $\Pi$ of 3 {\it projective} twistor components. From 
the full twistor structure of general solution (\ref{GENSOL})  gauge symmetry 
of the GSE (\ref{GSE}) of a rather peculiar type does follow, the so called 
"weak" one, when the gauge parameter is allowed to depend on coordinates 
only implicitly, i.e. via the spinor $\xi_{(0)}$ being transformed and its 
twistor counterpart $X\xi_{(0)}$. 

Now, in a usual way a Riemann metric $g$ of the Kerr-Shild type may be defined 
through the 1-form of the congruence $k=k_\mu dx^\mu$ as
\begin{equation}\label{RIMAN}
g = \eta + h~ k\otimes k ,
\end{equation}
with $\eta$ being the flat Minkowski metric. Well known is 
the fact that for a shear-free congruence the metric (\ref{RIMAN}) often satisfies 
the Einstein or Einstein-Maxwell (electro)vacuum equations under an appropriate 
choice of the "potential" function $h(x)$~\cite{ksmh}.  

Thus, at least the source-free Maxwell, Yang-Mills and the metric (effective 
gravitational) fields can be naturally assosiated with the solutions of the 
GSE (\ref{GSE}). Wonderfully, singularities of strengths (of curvatures) of 
all these fields coincide in space and time, being determined by a single 
condition 
\begin{equation}\label{SINGUL}
\det \Vert \frac{d \Pi^{(C)}}{d\xi_A} \Vert = 0,
\end{equation}
which specifies the points where the solution to (\ref{GENSOL}) becomes not 
unique. Geometrically, these points constitute a {\it caustic} of a correspondent 
shear-free light bundle. 
We'll see that the singular set, as defined by (\ref{SINGUL}), may be point-like, 
string-like and even a two-dimensional surface. In case when this set is bounded 
in 3-space we can identify it with a {\it particle-like} object whose time evolution 
is fully  governed by the field equations (\ref{GSE}) and may be obtained 
from the algebraic system (\ref{GENSOL}),(\ref{SINGUL}). 

These singular objects manifest, at a purely classical level,   
some properties of the real quantum particles; in particular, the value of 
electric charge $q$ is either zero or a whole multiple of the charge of the 
fundamental static solution; the latter corresponds just to the Kerr shear-free 
congruence with the ring-like singularity. The assosiated metric and  
electromagnetic field are identical to those of the Kerr-Newman solution in GTR, 
exept for the restrictions on the admissible value of electric charge    
~\footnote{In the dimensionless units we use; the numerical value 
itself is of no significance.} $q_0= \pm 1/4$ so that the quantity $q_0$ can 
be naturally identified with the {\it elementary charge}. In account of the 
well known property of the Kerr-Newman ansatz to have anomalous Dirac value of 
the {\it gyromagnetic ratio}~\cite{carter} we conclude that the fundamental static 
solution to the GSE reproduces all of the quantum numbers of the real 
electron. 

Throughout the whole  paper, we make no assumptions of either physical or 
mathematical character. Nor we introduce some additional dynamical equations 
for the fields involved, or use some Lagrangian structure. What we actually 
do is we study the intrinsic mathematical properties of the  
"generalized Cauchy-Riemann equations" as they are (precisely, a particular 
subclass of their solutions, represented by GSE (\ref{GSE})). 
Attempts to use such structures in theoretical physics have been repeatedly 
undertaken, especially in the works of F.G\"ursey and his 
co-workers 
~\cite{gt,ego} where the self-dual gauge and the chiral structures 
have been reformulated on the basis of quaternionic calculus. Most of considerations,  
however, made use of the only recognized version of quaternionic 
analiticity constructed by R.Fueter~\cite{fuet}. Unfortunately, his approach leads 
to trivial linear $\mathbb{C}$-like structure of the differentiability 
conditions ignoring the noncommutativity of the quaternion-type algebras  
and considerably limiting possible physical applications of the theory. 
 
 In the paper we don't intend to discuss the problem of noncommutative analysis 
in detail,  for this referring the reader e.g. to the profound paper of A.Sudbery 
~\cite{sud}. In the appendix we only frmulate the conditions of $\mathbb{B}$-
differentiability in the version earlier proposed in ~\cite {kass1,kass2,kass3,
kass5} and derive the GSE (\ref{GSE}) as a particular case of them. To our 
knowledge,  it is the only approach in which just the nonlinear structure of the 
generalized CR-equations naturally emerges, as a consequence of noncommutativity 
of the basic algebra (the situation well known for the gauge groups but quite 
unfamiliar in the analysis).   
In view of this, one should not be surprised that the field dynamics 
induced by the nonlinear differentiability conditions appears to be quite 
nontrivial and, perhaps, has something to do with real dynamics of particles. 
Anyway, the main goal of this article is to develop  
an unexpectedly rich and self-consistent "virtual" physics which is contained 
in a concise and invariant structure of the $\mathbb{B}$-differentiability 
conditions (\ref{GSE}) themselves, despite any additional phenomenological 
assumptions. 

The structures which arise in this context have some features in 
common with the I.Robinson's theorem~\cite{rob} relating shear-free congruences to 
(null) electromagnetic fields; with the approach of R.Penrose to the theory of 
spin-3/2 field relating twistor space to the charge space of electromagnetic 
field asssociated ~\cite{twcharg} and, especially, with spinor connections 
introduced by K.P.Tod ~\cite{Tod}.  From the generic viewpoint, our approach is 
adjacent to that  of Yu.Manin and G.Henkin ~\cite{manin} in which the nonlinear 
Dirac-Maxwell system is reduced to the linear Cauchy-Riemann equations in twistor 
space. On the whole, however, our approach does not appeal to any others and  
new physical concepts arising therein are not made up or artificial but 
quite inevitable in a consecutive mathematical scheme used.
  
Let us sketch out the organization of the article. In the second
section we establish the relativistic invariance of the GSE and obtain the 
4-eikonal equation for the components of the 2-spinor. In section 3 we 
demonstrate the functional dependence of the components of a twistor assosiated with 
the GSE and thus reduce the latter to the system of {\it algebraic} equations 
(\ref{GENSOL}). The fourth section is devoted to the analysis of the gauge symmetry 
of the GSE and to its specific relation to the projective transformations in the 
associated twistor space. In section 5 we study the self-dual structure of the 
GSE which follows from its integrability conditions and guarantees that the 
source-free Maxwell and Yang-Mills type equations are satisfied on the solutions 
of the GSE. In section 6 we establish the relations of GSE to the shear-free 
geodesic congruences and to the Riemann metrics of the Kerr-Schild type. The  
review of the "particle-like" solutions of GSE and the property of electric 
charge quantization are presented in section 7. In the last section 8 we conclude 
by the discussion of general problems and perspectives of the theory. In the 
appendix we formulate the conditions of differentiability for $\mathbb{A}$-valued 
functions of a noncommutative algebraic variable $Z\in \mathbb{A}$ and, for 
$\mathbb{A} \cong \mathbb{B}$, derive the GSE(1) as a particular case of 
these conditions.

Throughout the paper (exept in the appendix) the standard two-component spinor 
formalism is used. Upper case Latin indices range and sum over zero and one, 
and are raised and lowered by the symplectic spinors $\epsilon^{AB}$ and 
$\epsilon_{AB}$ respectively, as in ~\cite{pr}. By $\nabla_{AA^\prime}$ the 
usual spinor derivative operator in Minkowski space is denoted.

\section{Spinor structure and the eikonal equation}
\label{pbccv}

In the matrix representation of the biquaternion algebra $\mathbb{B}$ 
used in (\ref{GSE}) we regard $\xi$  as a column ($2\times 1$ matrix over 
$\mathbb{C}$), $\Phi$ -- as a 
$2\times 2$ complex matrix of general type, and $X$ -- as a hermitian 
matrix representing the coordinates of Minkowski space-time. System (\ref{GSE}) 
is evidently invariant under the global {\it Lorentz transformations} of coordinates
\begin{equation}\label{symm}
X\rightarrow L X L^+,~~~\xi\rightarrow \bar L^+\xi,~~~\Phi\rightarrow 
\bar L^+ \Phi \bar L,~~~\bar \Phi \rightarrow L \bar \Phi L^+,
\end{equation}
where $L\in SL(2,\mathbb{C})$ and $\bar \Phi$ denotes the matrix 
{\it adjoint} to $\Phi$: $\bar \Phi \Phi = (\det~\Phi)$ (we admit here 
the possibility for $\det \Phi$ to be zero). We see that the field $\xi(x)$ 
transforms as a (conjugate) 2-spinor whereas the components of the field 
$\bar \Phi(x)$ constitute a complex Lorentz 4-vector (we'll identify it later 
with the electromagnetic 4-potential).

In view of the form of transformations (\ref{symm}), functions $\xi,\Phi$ are 
mapped respectively to matrices $\xi_{A^\prime}$ and 
$\Phi_{B^\prime A}$, whereas $dX$ -- to a hermitian matrix $d{X^{AA^\prime}}$, 
with (un)primed indices $A,...A^\prime,...=0,1;0^\prime,1^\prime$ having usual 
2-spinor sense.
In matrix notation, (\ref{GSE}) has then the form
\begin{equation}\label{mGSE}
d\xi_{B^\prime}=\Phi_{B^\prime A}dX^{AA^\prime}\xi_{A^\prime},
\end{equation}
equivalent to a system of eight PDE 
\begin{equation}\label{delx}
\nabla_{AA^\prime} \xi_{B^\prime}=\Phi_{B^\prime A}\xi_{A^\prime}.
\end{equation}

Throughout the paper, we assume for both spinor components $\xi_{A^\prime}$ 
to be nonzero in the region of spacetime considered (otherwise, the solution 
to (\ref{GSE}) can be proved to be degenerate in some sense, with zero 
electromagnetic and other fields associated). We also consider 
all the functions $\{\xi_{A^\prime}(x),~\Phi_{B^\prime A}(x)\}$ to be analytical 
everywhere exept maybe at some subset of zero measure where they are allowed to 
turn to infinity.

Some properties of the solution $\xi(x)$ can be inferred directly from 
(\ref{delx}). Using the orthogonality identity $\xi^{A\prime}
\xi_{A^\prime}=0$ one easily find 
\begin{equation}\label{eikxx}
\nabla^{AA^\prime}\xi_{C^\prime} \nabla_{AA^\prime}\xi_{B^\prime}=0,
\end{equation}
which in particular implies the {\it eikonal} equation for any
function $\lambda(\xi_{A^\prime})$ of spinor components
\begin{equation}\label{eikx}
\nabla^{A A^\prime}\lambda(\xi_{A^\prime}) \nabla_{A A^\prime}\lambda(\xi_{A^\prime})=0.
\end{equation}

Returning to system (\ref{delx}), multiplying it by $\xi^{A^\prime}$ and again 
taking into account  $\xi^{A^\prime}\xi_{A^\prime}=0$ we reduce it to the 
following form~\cite{kr1}:
\begin{equation}\label{rbccv}
\xi^{A^\prime}\nabla_{AA^\prime}\xi_{B^\prime} = 0,
\end{equation}
where the 4-vector field $\Phi_{B^\prime A}$ has been dispensed with.
The latter may be recovered by using (\ref{delx}) with equal indices $A^\prime = 
B^\prime$~\footnote{We recall that both components of the 
spinor considered are assumed to be nonzero: otherwise all of the strengths 
(curvatures) vanish identically.}:
\begin{equation}\label{4pot}
\Phi_{A^\prime A}=\nabla_{A A^\prime} \ln \xi_{A^\prime};
\end{equation} 
certainly, no summation over $A^\prime$ is assumed in (\ref{4pot}).

Conversely, from (\ref{rbccv}) in view of a well-known property of 2-spinors 
it follows 
\begin{equation}\label{delxx}
\nabla_{AA^\prime}\xi_{B^\prime}=\varphi_{B^\prime A}\xi_{A^\prime},
\end{equation}
with some spintensor $\varphi_{B^\prime A}$. In compare of (\ref{delxx}) and (\ref{delx}) 
we conclude with the equivalence of (\ref{rbccv}) to the 
original spinor system (\ref{delx}) and, therefore, -- to the GSE (\ref{GSE}).

\section{Twistor structure and general solution of GSE}
\label{twist}

We turn now to the solutions of (\ref{rbccv}) or of GSE (\ref{delx}) 
equivalently. Remarkably, they happen to be completely determined by a 
{\it twistor structure} which can be naturally assosiated with the system 
considered. To demonstrate this, let us introduce another 2-spinor $\tau^{A}$ 
related to $\xi_{A^\prime}$ via the {\it Klein-Penrose correspondence} 
\begin{equation}\label{tv}
\tau^{A}= X^{AA^\prime}\xi_{A^\prime}.
\end{equation}
Then a pair of 2-spinors $T^a = (\xi_{A^\prime},\tau^A),~ a=0,1,0^\prime,1^\prime$, 
constitute a (null) twistor {\it incident} with a (real) Minkowski space-time 
point represented by $X^{A A^\prime}$. 
\footnote{For our purposes we may ignore imaginary unit $i$ in definition 
(\ref{tv}).}

According to definition (\ref{tv}) the wedge product of the differentials 
$d\xi_{A^\prime}$ and $d\tau^{A}$ may be formed as 
\begin{equation}\label{dtdx}
d \tau^{A} \wedge d \xi_{B^\prime} = X^{AA^\prime}d \xi_{A^\prime} \wedge 
d\xi_{B^\prime} +  \xi_{A^\prime}dX^{AA^\prime} \wedge d \xi_{B^\prime} .
\end{equation}
From (\ref{dtdx}) a rather obvious property about (nontrivial) twistors  
immediately follows: {\it at least} some two components of a generic twistor 
$T^a$ should be functionally independent (as functions of coordinates 
$X^{AA^\prime}$). Indeed, assuming conversely for all 
exterior products of the differentials $d\xi_{A^\prime}$ and $d\tau^{A}$ to 
vanish due to their functional dependence, we get from (\ref{dtdx}):
either both $\xi_{A^\prime}\equiv 0$ or $\nabla_{AA^\prime}\xi_{B^\prime}=0$, 
but in the second case both $\xi_{A^\prime}$ are constant and the two components 
of the spinor $\tau^A$ (\ref{tv}) are then evidently independent.    

If we subject now the spinor $\xi_{A^\prime}$ to the dynamical system 
(\ref{rbccv}) the remaining two components of the twistor {\it should depend} 
on the first independent two. Precisely, we intend to prove the following:
\vskip3mm \noindent{\bf Preposition.1}~{\it Iff $\xi_{A^\prime}$ is a
solution of (\ref{rbccv}) then the corresponding twistor $T^a$ has
two and only two functionally independent components}. \vskip1mm

Going to differentials in (\ref{tv}) and using (\ref{mGSE}), we come to equations 
for $d\tau^{A}$, which together with (\ref{mGSE}) themselves constitute a system of 
four equations
\begin{gather}
d \xi_{B^\prime} = \Phi_{B^\prime A} w^{A},\label{dxw}\\ d\tau^{B} = 
X^{BB^\prime} \Phi_{B^\prime A}w^{A}
+w^{B},\label{dtw}
\end{gather}
where the 1-forms $w^{A}=dX^{AA^\prime}\xi_{A^\prime}$ have been
introduced. Since the differentials of twistor components are
linear functions of the two 1-forms $w^{A}$ only, it becomes
obvious that the exterior product of any three is zero, resulting
in the desired functional dependence.

The same conclusion  could be reached in a slightly different way. Since
two twistor components are certainly functionally independent, we
can always find two equations in (\ref{dxw}),(\ref{dtw})
which allow to express $w^{A}$ through the differentials of
the corresponding twistor components. Substituting the resulting
expressions in remaining two equations, we end up with two
relations each containing three of the four differentials $d\xi_{A^\prime}$
and $d \tau^{A}$. These relations imply the sought-for  
functional dependence between any three twistor components, a fact
that can be expressed in a more symmetric form~\cite{kr1}
\begin{equation}\label{sol}
\Pi^{(C)}(T^a)\equiv \Pi^{(C)}(\xi_{A^\prime},\tau^B)=0,
\end{equation}
where $\{\Pi^{(C)}\},~C=0,1$ are two arbitrary but independent holomorphic 
functions of four complex variables.

Conversely, the algebraic system (\ref{sol}) implicitly determines
$\xi_{A^\prime}$ and it is easy to check, by differentiating
(\ref{sol}) and multiplying by $\xi^{A^\prime}$, that
$\xi_{B^\prime}$ satisfies the system
\begin{equation}\label{diffpi}
\frac{d
\Pi^{(C)}}{d\xi_{B^\prime}}\xi^{A^\prime}\nabla_{AA^\prime}\xi_{B^\prime}=
=0,
\end{equation}
which is equivalent to (\ref{rbccv}) except at some singular
points (see below). Successively resolving system (\ref{sol}) at each 
space-time point $X^{A A^\prime}$ with respect to  $\xi_{A^\prime}$ and 
substituting the resulting solution in (\ref{4pot}) to find the corresponding 
"potentials" $\Phi_{B^\prime A}$ we obtain a solution to the GSE starting 
from the algebraic constrains (\ref{sol}). This furnishes the proof of 
preposition.1. $\blacksquare$
   
Thus, algebraic system of equations (\ref{sol}) 
implicitly determines the {\it general (analytical) solution} $\{\xi_{A^\prime}, 
\Phi_{B^\prime A}\}$ of the GSE. Points where the equations (\ref{sol}) have 
multiple roots, i.e. cannot be {\it in a unique way} resolved for 
$\xi_{A^\prime}$ satisfy according to (\ref{diffpi}) the equation
\begin{equation}\label{sin}
\det\Vert\frac{d \Pi^{(C)}}{d\xi_{A^\prime}}\Vert =0.
\end{equation}
These points constitute a singular set for electromagnetic field which 
in the next section will be assosiated with the quantities $\Phi_{A^\prime A}$. 
Together with (\ref{sol}) the last equation allows us to determine the
shape and the time evolution of singularities (see below, section 7).

Geometrically, the algebraic system (\ref{sol}) defines a two-dimensional complex 
surface in the twistor space $\mathbb{C}^4$ (precisely, in the subspace of 
null twistors). Points of intersection of this surface 
with two-dimensional planes formed by (null) twistors (\ref{tv}) represent 
the solution $\xi_{A^\prime}$ to the GSE ({\it multivalued} in general) for each 
fixed space-time point $X^{A A^\prime}$. Singularities are then the pre-images 
(in $M$) of the points of twistor space at which the planes (\ref{tv}) are  
{\it tangent} to the surfaces (\ref{sol}), so that the singular set will be the 
same for all of the branches of a multivalued solution. Note that we ignore here 
the generally considered projective structure of the twistors which,  
in the framework of this approach, has some peculiarities and is related to an 
exotic gauge symmetry of GSE. These issues are discussed in the next section.

\section{Projective transformations of twistors and "weak" 
gauge symmetry of the GSE}

For an appropriate electrodynamical interpretation we need to 
establish the gauge invariance of the GSE, which will be dealt with in this 
section. Specifically, we shall study the symmetry of (\ref{delx})
under transformations
\begin{equation}\label{gauge}
\xi_{A^\prime} \to \xi^\prime_{A^\prime}=\alpha(x)\xi_{A^\prime},
\end{equation}
where $\alpha(x)$ is a smooth complex function of coordinates. Using equations 
(\ref{rbccv}), it's readily seen that $\alpha$ cannot be an
arbitrary function of coordinates, it rather satisfies the equation
\begin{equation}\label{alph1}
\xi^{A^\prime}\nabla_{AA^\prime}\alpha=0,
\end{equation}
from which in view of 2-spinors' properties follows 
\begin{equation}
\nabla_{AA^\prime} \alpha = \rho_A \xi_{A^\prime}
\end{equation}
for some $\rho_A$ and, consequently, -- the eikonal equation for $\alpha(x)$
\begin{equation}\label{eikalph}
\nabla^{AA^\prime}\alpha \nabla_{AA^\prime}\alpha = 0.
\end{equation}

Before we carry on, we need to establish an auxiliary result. Let
us rewrite equation (\ref{dtw}) in partial derivatives
\begin{equation}\label{delt}
\nabla_{AA^\prime}\tau^{B} = X^{BB^\prime}
\Phi_{B^\prime A}\xi_{A^\prime} +\xi_{A^\prime}\delta^{B}_{A}.
\end{equation}
Using the orthogonality $\xi^{A^\prime}\xi_{A^\prime}=0$, we immediately verify 
the validity of the following relations:
\begin{gather}
\nabla^{AA^\prime}\xi^{B^\prime}\nabla_{AA^\prime}\tau^{B}=0,\label{eikxt}\\
\nabla^{AA^\prime}\tau^{B}\nabla_{AA^\prime}\tau^{C}=0,\label{eiktt}
\end{gather}
which along with equation (\ref{eikxx}) lead to the eikonal
equation for any function $\lambda(T^a)$ of twistor components
\begin{equation}\label{eikonalt}
\nabla^{AA^\prime} \lambda(T^a)\nabla_{AA^\prime} \lambda(T^a)=0.
\end{equation}

Going now back to the main goal of this section and taking into
consideration the eikonal equation for $\alpha(x)$, we are guided
by the equation (\ref{eikonalt}) to conjecture the following
\vskip3mm \noindent{\bf Preposition.2}~{\it Transformation of the
type (\ref{gauge}) are symmetries of (\ref{delx}) iff $\alpha$ is a
function of $T^a$ and $\Phi_{B^\prime A}$ transforms according to}
\begin{equation}\label{gradpot}
\Phi_{B^\prime A} \to \Phi^\prime_{B^\prime A} = \Phi_{B^\prime A} +
\nabla_{A B^\prime} \ln \alpha.
\end{equation}

Replacing in (\ref{delx}) $\xi_{A^\prime}$ and $\Phi_{B^\prime A}$ by their 
transformed values, after some simplification we obtain the following 
condition of form-invariance of (\ref{delx}): 
\begin{equation}\label{alph2}
\xi_{B^\prime}\nabla_{A A^\prime}\alpha - \xi_{A^\prime}\nabla_{A B^\prime}\alpha = 0,
\end{equation}
which is skew-symmetric in $A^\prime, B^\prime$ and therefore equivalent to the 
equation (\ref{alph1}) for $\alpha(x)$. 
Taking now into account equations (\ref{delx}) and (\ref{delt}) and
carrying out simple manipulations we show that if $\alpha=\alpha(T^a)=\alpha
(\xi_{A^\prime},\tau^B)$ then equation (\ref{alph1}) is identically satisfied. 
This proves the sufficient part of the preposition.

To prove the converse, suppose that transformation (\ref{gauge})
is a symmetry of (\ref{delx}). This yields the following:
\begin{equation}\label{dalph}
\xi_{B^\prime} d\alpha = \alpha(\Phi^\prime_{B^\prime A}-
\Phi_{B^\prime A}) w^A,
\end{equation}
where, as before, $w^{A}=dX^{AA^\prime}\xi_{A^\prime}$. Making use
of (\ref{dxw}) and (\ref{dtw}), it's easy to see that the
exterior product of equation (\ref{dalph}) with any two
differentials of the twistor components vanishes, leading to the
functional dependence of $\alpha$ on the corresponding twistor
components. More symmetrically, this result can be expressed as
asserted in the preposition, i.e. $\alpha=\alpha(T^a)$.
Passing then to partial derivatives in (\ref{dalph}) we obtain the relation 
\begin{equation}\label{delalph}
\xi_{B^\prime}\nabla_{AA^\prime}\ln \alpha = (\Phi^\prime_{B^\prime A}-
\Phi_{B^\prime A})\xi_{A^\prime}~,
\end{equation}
from which the desired transformation rule (\ref{gradpot}) for the "potentials" 
$\Phi_{B^\prime A}$
follows (in a special case when the primed subscripts
are equal). This completes the proof of preposition.2. $\blacksquare$

Some words are in order about the nature of transformations
\begin{equation}\label{rgauge}
\xi \to \xi^{\prime} = \alpha(T^a)\xi.
\end{equation}
which may be called {\it restricted}, or {\it weak} gauge transformations 
~\cite{kr2}. We remark that according to its definition (\ref{tv}) the conjugate 
spinor $\tau^B$ transforms in a similar way
\begin{equation}\label{taugauge}
\tau \to \tau^\prime=\alpha(T^a)\tau
\end{equation}
so that both (\ref{rgauge}),(\ref{taugauge}) together imply that the gauge 
symmetries of GSE may be considered as the transformations of twistors of the form
\begin{equation}\label{tvgauge}
T \to T^\prime = \alpha(T^a) T .
\end{equation}

It's clear that  composition of transformations (\ref{tvgauge}) is a 
transformation of the same type. Their associativity is also obvious. But the 
existence of inverse transformation is not so evident. However, according to
preposition.1, $T^a$ and its image $T^{\prime a}$ both have only
two functionally independent components, and $\alpha(T^a)$
depends, essentially, on these two components (of $T^a$). So we
can always express the two independent components of $T^a$ through
those of $T^{\prime a}$. Substituting them in $\alpha^{-1}(T^a)$ results in 
the inverse transformation $T=\alpha^{-1}(T^{\prime a})T^\prime$ which 
is of the same type as (\ref{tvgauge}). Hence {\it these
transformations constitute a group}! In fact it is a proper
subgroup of the full $\mathbb{C}(1)$-gauge group of transformations 
(\ref{gauge}), the latter itself being generally not a symmetry of the GSE . 
The statement that this subgroup is a proper one, becomes quite evident if we recall that
$\alpha(T^a)$ should be subject to the eikonal equation (\ref{eikalph}).  

Finally, we note that under the transformations (\ref{rgauge}) the {\it trace} part 
of the matrix 1-form $A \equiv Tr(\Phi dz) = \Phi_{A^\prime A} dX^{A A^\prime}$
with components $\bar \Phi_{A A^\prime}\equiv\Phi_{A^\prime A}$ transforms gradient-wise
\begin{equation}\label{potgauge}
A \to A + d\ln \alpha,
\end{equation}
as the electromagnetic potential 1-form does under the gauge transformations 
(this may be also seen from the expression (\ref{4pot})). In view of the 
4-vector properties of $\bar \Phi$ under Lorentz transformations (\ref{symm}) we 
are brought to adopt the interpretation of the 1-form $A$ as the potentials and of the 
respective gauge-invariant 2-form $F=dA$ as the electromagnetic field strengths 
(of course, up to an arbitrary scale factor only). In the following section we 
obtain Maxwell equations for this 2-form, therefore elaborating its 
electromagnetic interpretation.

Let us now look at the gauge transformations (\ref{tvgauge}) from the 
viewpoint of the geometry of twistor space. The Abelian nature of the transformations 
studied results in the fact that the {\it ratio} of any two twistor components 
$T^a$ is  invariant under them. Thus, such transformations are {\it projective} 
in origin. Not only the planes (\ref{tv}) but also the surfaces (\ref{sol}) are 
form-invariant under transformations (\ref{tvgauge}) and, consequently,  
give rise to  another solution of the GSE (with the same electromagnetic 2-form 
$F$). So we may consider the {\it equivalevce classes} of the solutions (and of 
the surfaces (\ref{sol}) respectively) which can be 
obtained one from another via the gauge transformations (\ref{tvgauge}). 
That's why we may restrict ourselves to consider only  
the {\it projective twistor space} $CP^3$. However, projective structure 
of this type differs essentially from the conventional one which originates 
from the transformations of the full gauge group (\ref{gauge}). We shall return 
to this problem in section 6, and for the time being shall deal with the full 
structure of the space of (null) twistors. 

\section{Integrability conditions, self-duality and the source-free gauge 
equations}\label{int}
As was mentioned in the introduction, the GSE (\ref{GSE}) may be viewed at as 
the condition that must be met for a spinor $\xi(x)$ to be covariantly 
constant with respect to the $\mathbb{B}$-connection 1-form 
\begin{equation}\label{conn}
\Omega=\Phi dz .
\end{equation}
It may be interesting to note that in the 4-vector representation 
$\mathbb{B}$-connection (\ref{conn}) turns into the affine connection of the 
form~\cite{kass1,kass2} 
\begin{equation}\label{weyl}
\Gamma_{\nu\rho}^\mu = \delta_\nu^\mu \Phi_\rho + \delta_\rho^\mu \Phi_\nu - 
\eta_{\rho\nu}\Phi^\mu - i\epsilon^\mu_{.\nu\rho\lambda}\Phi^\lambda
\end{equation}
which gives rise to the effective complexified geometry of {\it Weyl-Cartan type}.  
For this $\mathbb{B}$-induced geometry the Weyl vector of nonmetricity and the 
pseudotrace of the torsion tensor appear to be proportional to each other and 
are both expressed via the components of the basic gauge field $\Phi(x)$~\footnote
{Vector fields covariantly constant with respect to the torsion-free Weyl 
connection have been studied in ~\cite{kr3}; they are closely related to the 
symmetries of Weyl manifolds~\cite{hall}. Relations between the nonmetrical 
and torsion parts of such connections were considered in ~\cite{Tod}.}. 

According to definition (\ref{conn}), the initial GSE (\ref{GSE}) may be 
rewritten as follows:
\begin{equation}\label{ccv}
d \xi=\Omega \xi
\end{equation}
{\it The GSE is overdetermined} (8 equations for 6 unknown 
functions) and both the spinor and the "connection" gauge fields 
are to be determined from it. Dynamics of the connection field $\Omega(x)$ can  
be obtained by external differentiation of (\ref{ccv}) which yields
\begin{equation}\label{compat}
R \xi \equiv (d \Omega - \Omega \wedge \Omega) \xi = 0,
\end{equation}
where in parentheses the matrix {\it curvature 2-form} $R$ of the connection    
(\ref{conn}) arises. Since the spinor $\xi$ is not arbitrary but subject to 
(\ref{ccv}) the {\it integrability conditions} (\ref{compat}) don't imply 
the triviality of curvature~\footnote{At this point our approach differs 
essentially from that of Buchdahl~\cite{buch}, Penrose~\cite{prlast} and 
Plebanski~\cite{pleb} who assumed that the integrability conditions like (\ref{compat}) should 
be satisfied identically for an arbitrary spinor field, in order to ensure the 
existence of an "exact set" of solutions to field equations.},  
instead they lead to its {\it self-duality}~\cite{kass1,kass2}.     
                           
To demonstrate this, we note that for connection ({\ref{conn}) 
the curvature $R$ is of the following, rather specific form
\begin{equation}\label{curv}
R = (d \Phi - \Phi dz \Phi) \wedge dz \equiv \pi \wedge dz ,
\end{equation}
where a new $\mathbb{B}$-valued 1-form $\pi$ emerges, with the components 
\begin{equation}\label{compon}
\pi_{A^\prime C}=\pi_{A^\prime C B B^\prime}dX^{BB^\prime}=
(\nabla_{BB^\prime}\Phi_{A^\prime C}-\Phi_{A^\prime B}\Phi_{B^\prime C})
dX^{BB^\prime}.
\end{equation}
The integrability conditions (\ref{compat}) take then the form 
$(\pi \wedge dz) \xi =0$ or, in matrix notation  
\begin{equation} 
\pi_{A^\prime C B B^\prime} dX^{B B^\prime} \wedge dX^{C C^\prime} 
\xi_{C^\prime} =0 .
\end{equation}
Making use of symmetry properties we obtain from the last relation
\begin{equation}
\pi_{A^\prime ~ C(B^\prime}^{~~C}\xi_{C^\prime)} = 0,
\end{equation}
so that for any nontrivial solution $\xi(x)$ it follows
\begin{equation}\label{compi}
\pi_{A^\prime~CB^\prime}^{~~C}\equiv \nabla_{C B^\prime}\Phi_{A^\prime}^{~~C}+
\Phi_{B^\prime C}\Phi_{A^\prime}^{~~C} = 0.
\end{equation}
  
Decomposing now in a usual way the curvature (\ref{curv}) into the self- and 
antiself-dual parts it is easy to verify that equations (\ref{compi}) are 
just the conditions for its self-dual part to vanish, whereas the other 
(antiself-dual) one $\bar R$ takes the form
\begin{equation}\label{antiself}
\bar R_{A^\prime(BC)}^{~~C^\prime}=\nabla_{~(B}^{C^\prime}\Phi_{A^\prime C)} - 
\Phi_{~(B}^{C^\prime}\Phi_{A^\prime C)}
\end{equation}
and satisfy additional integrability conditions $\bar R \xi=0$ (we'll not make 
use of them below). 

Thus, though the curvature 2-form (\ref{curv}) of the connection 1-form 
(\ref{conn}) is not identically (anti)self-dual (i.e. self-dual in the "strong" sense), 
it nesessarily becomes (anti)self-dual {\it on the solutions of the GSE}. For 
this reason, the property has been called in~\cite{kr1} {\it weak self-duality}. 

Physically, the expression (\ref{antiself}) represents the strength of 
a matrix gauge field; in particular, its trace part $F_{BC}=\bar 
R_{A^\prime~~(BC)}^{~~A^\prime}=\nabla_{~(B}^{A^\prime}\Phi_{A^\prime C)}$ 
corresponds to the aforedefined electromagnetic field strength $F=dA$ whereas 
the trace-free part of (\ref{antiself}) defines the strength of a {\it Yang-Mills 
type' field}~\footnote{Owing to the restricted (weak) gauge symmetry this 
is not precisely what is usually understood as the YM fields; however, the 
dynamical structure of the gauge equations is completely the same, the 
restrictions are imposed only on the solutions.}

Indeed, in view of {\it Bianchi identities}
\begin{equation}\label{bianchi}
dR \equiv \Omega \wedge R - R \wedge \Omega, 
\end{equation}
self-duality of curvature $R+iR^* = 0$ leads to the source-free Maxwell
equations for the electromagnetic 2-form $F=Tr(R)=R_{A^\prime}^{~~A^\prime}$
\begin{equation}\label{maxeq}
dF^*=0= dF \equiv 0,
\end{equation}
and to the equations of Yang-Mills type for the trace-free part of curvature
${\bf F}_{A^\prime}^{~~B^\prime}=R_{A^\prime}^{~~B^\prime}-\frac{1}{2}
F\delta_{A^\prime}^{~B^\prime}$.

Generally speaking, electromagnetic 2-form $F$ is a $\mathbb{C}$-valued
field, yet in view of its self-duality it reduces to an
$\mathbb{R}$-valued 2-form $\tt{F}$ related to $F$ by
\begin{equation}\label{freal}
F=\tt{F}-i\tt{F}^*,
\end{equation}
for which Maxwell equations do hold as well, so that the number of its 
degrees of freedom is just equal to that of a usual real electromagnetic field. 
Explicitly, from the symmetric part of the integrability conditions (\ref{compi}) 
we get for the $\mathbb{C}$-valued "electric" $\vec E$ and "magnetic" $\vec H$ 
field strengths 
\begin{equation}\label{eh}
\vec E + i\vec H = 0,
\end{equation}
so that $\Im (\vec H) = \Re (\vec E)$, $\Im (\vec E) = - \Re (\vec H)$ and the 
pair $\{\Re (\vec E), \Re (\vec H)\}$ represents an $\mathbb{R}$-valued 
electromagnetic field subject to Maxwell equations. Note that from the 
skew-symmetric part of (\ref{compi}) in  the 4-vector form we get the 
following "inhomogeneous Lorentz condition"~\cite{kass1,kass2} for the 
$\mathbb{C}$-valued electromagnetic potentials $A_\mu=\Phi_{A^\prime A}$:
\begin{equation}\label{loren}
\partial_\mu A^\mu + 2 A_\mu A^\mu = 0,
\end{equation}
which is also identically satisfied on each solution to the GSE. Condition 
(\ref{loren}) is by no means gauge invariant in a usual sense but it {\it is} 
invariant with respect to the weak gauge transformations (\ref{rgauge}) 
provided the potentials satisfy the GSE.

As to the fields of Yang-Mills type, they may be always expressed via the 
electromagnetic field strengths and the spinor $\xi_{A^\prime}$ itself and 
for this reason can't be regarded independently. Note also that the real or 
imaginary part of the trace-free curvature ${\bf F}_{A\prime}^{~~B^\prime}$ 
being taken separately would not satisfy the source-free YM equations, in view 
of nonlinearity of the latters. Therefore, the YM-like fields here {\it are 
necessarily complex}. Other details about the pecularities of YM fields in the 
framework of algebrodynamical approach can be found in~\cite{kass2}.

\section{GSE and the null shear-free geodesic congruences}
\label{grav} Let us recall that via the elimination of potentials $\Phi_
{A^\prime A}$  the GSE (\ref{GSE}) takes the form (\ref{rbccv}). Once its  
solution $\xi_{A^\prime}(x)$ is found, a field of a null 4-vector $k_\mu(x),~ 
k_\mu k^\mu =0$ can be defined as
\begin{equation}\label{cong}
k=k_\mu dx^\mu \equiv \xi_A\xi_{A^\prime} dX^{AA^\prime}.
\end{equation}
Vector lines of this field define a null congruence for which the 
shear-free criterion~\cite{pr}   
\begin{equation}\label{sfc}
\xi^{A^\prime} \xi^{B^\prime} \nabla_{AA^\prime} \xi_{B^\prime}=0
\end{equation}
follows readily from (\ref{rbccv}). Hence each solution $\xi_{A^\prime}$ of 
the GSE actually defines a {\it null shear-free geodesic congruence}
(SFC). Contrary to (\ref{rbccv}), the SFC equations (\ref{sfc}) are invariant 
under the full complex Abelian gauge group (\ref{gauge}) and reduce to the 
system of two equations in partial derivatives 
\begin{equation}\label{gsol}
\nabla_w G=G\nabla_{u}G,~~~\nabla_{v}G=G\nabla_{\bar w}G,
\end{equation}
where $G=\xi_{1^\prime}/\xi_{0^\prime}$ is the gauge invariant, 
and the following generally accepted notation has been used:
\begin{equation}\label{zcoor}
X^{AA^\prime}=
  \begin{pmatrix}
    u & w \\
    \bar w & v
  \end{pmatrix}\equiv
  \begin{pmatrix}
    x^0+x^3  & x^1-ix^2 \\
    x^1+ix^2 & x^0-x^3
  \end{pmatrix}
\end{equation}
Four real quantities $\{x^i; x^0\},~i=1,2,3$ correspond to the Cartesian space and time 
coordinates respectively. Note that as for the individual spinor components 
$\xi_{A^\prime}$, they remain fully indeterminate by the SFC equations which 
impose restrictions only on their quotient $G(x)$. 

Let us compare this with the GSE (\ref{rbccv}). The latter is equivalent  
to the system of {\it four} equations for only two spinor compoments 
$\xi_{A^\prime}$
\begin{equation}\label{cbccv}
\nabla_{w}\xi_{A^\prime}=G\nabla_u \xi_{A^\prime}, ~~~ 
\nabla_v \xi_{A^\prime} = G\nabla_{\bar w} \xi_{A^\prime} 
\end{equation}
from which again the equations (\ref{gsol}) follow for the quotient $G(x)$. 
Multiple solutions of (\ref{cbccv}) with the same $G$ correspond to different 
potentials but have the same strengths of the electromagnetic and YM fields  
assosiated. In view of this, further on we identify the GSE and SFC equations 
by considering only the {\it projectively invariant part} of the GSE represented 
by system (\ref{gsol}) (one may regard this as a fixing of the 
gauge $\xi_{0^\prime}=1$).  

General analytical solution of (\ref{gsol}) for $G(x)$ immediately follows now 
from the preposition.1. in the form of an algebraic equation  
\begin{equation}\label{rsol}
\Pi(G, \tau^0, \tau^1)= \Pi(G,u+w G, \bar w+vG)=0,
\end{equation}
which determines implicitly the function $G(x)$. Here $\Pi$ is an 
arbitrary holomorphic function of three complex variables. Equation 
(\ref{rsol}) manifests the functional dependence of three components 
$G,\tau^0,\tau^1$ of a projective twistor $T^a$ assosiated with the 
solutions of GSE. For the SFC equations (\ref{sfc}) the equivalent result is 
well known as the {\it Kerr theorem}~\cite{pr}. Note that the solutions of (\ref{gsol}) 
in the form (\ref{rsol}) are defined exept at the points of the singular set 
whose equation (\ref{sin}) simplifies now to  
\begin{equation}\label{rsin}
P \equiv\frac{d\Pi}{dG}=0.
\end{equation}

By multiplying the two equations of (\ref{gsol}) we obtain once more the 
4-eikonal equation for $G(x)$ in the form
\begin{equation}\label{wafront}
\nabla_u G\nabla_v G -\nabla_w G \nabla_{\bar w} G = 0,
\end{equation}
while by differentiating them we verify that $G(x)$ satisfies also the linear 
d'Alembert equation 
\begin{equation}\label{wave}
\Box G(x) \equiv (\nabla_u\nabla_v - \nabla_w\nabla_{\bar w})G(x) = 0.
\end{equation}
Note that in view of (\ref{wafront}) every $C^2$-function $\lambda(G)$
is also harmonic on the solutions of the GSE, 
\begin{equation}\label{harm}
\Box \lambda(G) = 0.
\end{equation}

Using now ansatz (\ref{4pot}) for the potentials $\Phi_{A^\prime A}$ 
and taking into account (\ref{wafront}), we can 
express the electromagnetic field strengths via the 2-nd order derivatives 
of $\ln G$ as 
\begin{equation}\label{streng}
F_{00}=\nabla_u\nabla_{\bar w}\ln G,~~F_{11}=\nabla_v\nabla_w\ln G,~~
F_{01}=\nabla_w\nabla_{\bar w}\ln G ,
\end{equation}
so that satisfaction of the source-free Maxwell equations is then ensured 
for (\ref{streng}) in consequence of the wave equation (\ref{harm}) for $\lambda=
\ln G$. Differentiating twice the identity (\ref{rsol}) with respect to 
the (spinor) space-time coordinates we obtain finally for the strengths (\ref{streng}) 
the following symmetric expression:
\begin{equation}\label{streng2}
F_{AB}=\frac{1}{P}\left(\Pi_{AB}-\frac{d}{dG}(\frac{\Pi_A\Pi_B}{P})\right),
\end{equation}
where the function $P$ is defined by (\ref{rsin}) and $\{\Pi_A,\Pi_{AB}\},~ 
A,B=0,1$ denote the (1-st and 2-nd order) partial derivatives of $\Pi$ with 
respect to its twistor arguments $\tau^0,\tau^1$. We return to this expression 
below. 

Close relation of the GSE to the SFC makes it possible to introduce one more 
geometrophysical structure -- an effective {\it Riemann metric}.  In fact, 
it's well-known~\cite{dks,ksmh} that we can transform the flat Minkowski metric 
$\eta_{\mu\nu}$ into a metric $g_{\mu\nu}$ of the Kerr-Schild type 
\begin{equation}\label{metr}
g_{\mu\nu} = \eta_{\mu\nu} + hk_\mu k_\nu,
\end{equation}
and that the main characteristics of SFC $k_\mu$ (geodesity, twist and 
shear) are preserved under this transformation. Here $h$ is a scalar field 
and the congruence $k$ given by (\ref{cong}) acquires a projectively invariant form
\begin{equation}\label{cong2}
k=du+\bar G dw + G d\bar w + G\bar G dv ,
\end{equation}
$\bar G$ being complex conjugated of $G$.
Now we resort to the results of the classical paper~\cite{dks} where it has 
been proved that the metric (\ref{metr}) satisfies  Einstein-Maxwell 
electrovacuum system for any function $G$ obeying algebraic constraint 
(\ref{rsol}), with the function $\Pi$ {\it linear} in twistor arguments 
$\tau^0,\tau^1$~:
\begin{equation}\label{PiG}
\Pi = \varphi + (qG + s)\tau^1-(pG + \bar q)\tau^0.
\end{equation}
Here $\varphi=\varphi(G)$ is an arbitrary analytic function of
the complex variable $G$, $s$ and $p$ are real constants and $q$
is a complex constant. Not going into details we just note
that according to the results of~\cite{dks} the scalar field $h$
in (\ref{metr}) is determined, up to an arbitrary real
constant, by the function $\Pi$ and some another function $\Psi(G)$
independent of $\varphi(G)$ and related to the electromagnetic
field arising therein. These electromagnetic fields are defined in
the curved space-time with metric (\ref{metr}), and generally
they are different from those emerging in our approach and
satisfying Maxwell equations in the {\it flat} space-time\footnote{At the same time 
they both are generally different from the fields which may be defined for 
the SFC using the Penrose twistor transform, see~\cite{pr,pr2}.}. However, for
the most fundamental Reisner-N\"ordstrem and Kerr-Newman solutions these 
fields coincide, the only difference being in that in our approach the
electric charge is fixed in magnitude by the GSE itself (see the next section).

In~\cite{kw,dks} it was shown that singularities of Riemann curvature of 
the Kerr-Schild metric (\ref{metr}) correspond just to the condition (\ref{rsin}). 
On the other hand, the expression (\ref{streng2}) demonstrates that the same 
condition $P=0$  determines the set of singular points of the electromagnetic 
field. It may be verified that this is true also for the strengths of Yang-Mills 
fields assosiated with the solutions of the GSE.

Hence, with each solution of the GSE an electromagnetic,
a $\mathbb{C}$-valued Yang-Mills and an effective gravitational field can be 
naturally associated, their
singularities are all determined by equation (\ref{rsin}) and
therefore coincide in space and time. This makes it possible, in the framework of
the algebrodynamical approach based on the GSE, {\it to consider particles as 
common singularities of all fields involved}. This general concept will be 
developed in the next section.

\section{Quantization of electric charge and "particle-like" solutions of GSE}
\label{bound} We'll briefly review now the main solutions of the GSE known to 
date which all may be obtained by an appropriate choice of the function $\Pi$ 
and solving subsequently the algebraic equation (\ref{gsol}). In order  
to find the solutions in a simple explicit form one usually restricts the 
consideration to functions $\Pi$ which are {\it quadratic} in G (linear 
functions result in zero electromagnetic strengths (\ref{streng})). 
The fundamental {\it static} solution is generated by a function 
$$
\Pi = G\tau^0-\tau^1 + 2ia \equiv G(u+ w G)-(\bar w + vG)+ 2ia=0,
$$
($a=Const \in \mathbb{R}$), from where we get
\begin{equation}\label{kerr}
G = \frac{\bar w}{(z+ia)\pm r_*}\equiv\frac{x+iy}{(z+ia)\pm \sqrt{x^2+y^2+(z+ia)^2}}.
\end{equation}
Electromagnetic field (\ref{streng}) assosiated with this solution 
\begin{equation}\label{elctrfd}
\vec E - i\vec H = \pm\frac{\vec r_*}{4(r_*)^{3/2}};~~~~~~~~(\vec E + i\vec H = 0),
\end{equation}
where $\vec r_* =\{x,y,z+ia\}$ has a {\it ring singularity} of radius $a$, an 
only possible electric charge $q=\pm 1/4$ (in the dimensionless units we use), 
a dipole magnetic and a quadrupole electric moments equals to $qa$ and $qa^2$ 
respectively~\cite{kr2}. Apart from the restriction on charge, 
the electromagnetic field (\ref{elctrfd}) together with 
the Riemann metric assosiated with (\ref{kerr}) via the SFC (\ref{cong2}),  
accurately reproduce the field and metric of the Kerr-Newman solution (in the 
coordinates used in~\cite{dks}). Particularly, for $a=0$ the solution (\ref{kerr}) 
represents the {\it stereographic map} $S^2\to \mathbb{C}$ whereas the fields 
turn into the Coulomb one and the Reisner-N\"ordstrem metric respectively.
 
Quantization of electric charge seems to be a profound property of the GSE-solutions 
discovered in~\cite{kass1,kass2}. It is a consequence of self-duality 
condition (\ref{eh}) which together with the gauge symmetry of GSE ensures the 
relation $q=N/4, N\in \mathbb{Z}$ for 
the values of electric charge assosiated with every solution of the GSE. 
This property has both topological and dynamical origins, the latter 
being related to the overdetermined structure of the GSE. The proof of the 
general theorem on charge quantization will be presented elsewhere. Contrary to the 
recently developed~\cite {ranada,zhuravl} approaches to the problem of 
quantization of electric charge, which are purely topological, in the 
framework of the GSE the charge of the  
fundamental static solution (\ref{kerr}) can be of only one fixed and minimal 
possible value and, therefore, can be naturally identified with the  
{\it elementary charge}. Together with the well-known property of Kerr-Newman 
solution to fix the gyromagnetic ratio $g=2$ equal to the ratio 
of the Dirac particle~\cite{carter} the appearance of elementary electric 
charge within the theory makes it much more legitimate to interprete the 
fundamental solution (\ref{kerr}) as a classical model of electron (
in comparison, say, with the models of Lopez~\cite{lopes}, Israel~\cite{israel} 
and Burinskii~\cite{burin1} based on Einstein-Maxwell theory itself).
 
According to a general theorem proved in~\cite{kw} {\it static} solutions 
of the SFC (and thus of the GSE) for which the singular set is bounded in 3-space (below we call them 
{\it particle-like}~\cite{kt}) are all exhausted by 
the Kerr solution (\ref{kerr}) (up to 3-translations and 
3-rotations). If, however, we remove the condition for a solution to be static 
and get out of the class of functions (\ref{PiG}) dealt with in~\cite{dks} we 
discover a lot of time-dependent "particle-like" solutions with bounded 
singularities of different dimensions, 3-shapes and time evolution. 
   
In particular, an {\it axisymmetric} solution of a pacticle-like type generated 
by the function 
\begin{equation}\label{jozeph}
\Pi = \tau^0\tau^1 + b^2 G^2 = 0, ~~~b=Const,
\end{equation}
has been found in~\cite{kr1,diss}. For real $b$ it corresponds to the 
case of two singularities with elementary charges $+1/4$ and $-1/4$ undergoing 
head-on hyperbolic motion for which the electromagnetic field 
\begin{equation}\label{fieldborn}
E_\rho=\pm\frac{8b^2\rho z}{\Delta^{3/2}},~~~E_z = \mp\frac{4b^2 M}{\Delta^{3/2}},~~~
H_\varphi =\pm\frac{8b^2\rho t}{\Delta^{3/2}},
\end{equation}
is identical to that of the {\it Born solution}. Here the following notation  
is used: 
$$\rho^2 = x^2+y^2,~~~s^2 = t^2-z^2,~~~M=s^2+\rho^2+b^2,~~~\Delta=M^2-4s^2\rho^2,$$ 
and singularities are defined by the condition $\Delta=0$. 
For imaginary $b=ia,~ a\in \mathbb{R}$ one has at $t=0$ a neutral ring-like 
singularity which then expands to a {\it torus}. After an interval of time 
$t>\vert a \vert$ singularity turns into a {\it self-intersecting torus} 
depicted in Fig.1.  

\begin{center}
\begin{figure}[ht]
~~~~~~~~~~~~~~~~~
\epsfig{file=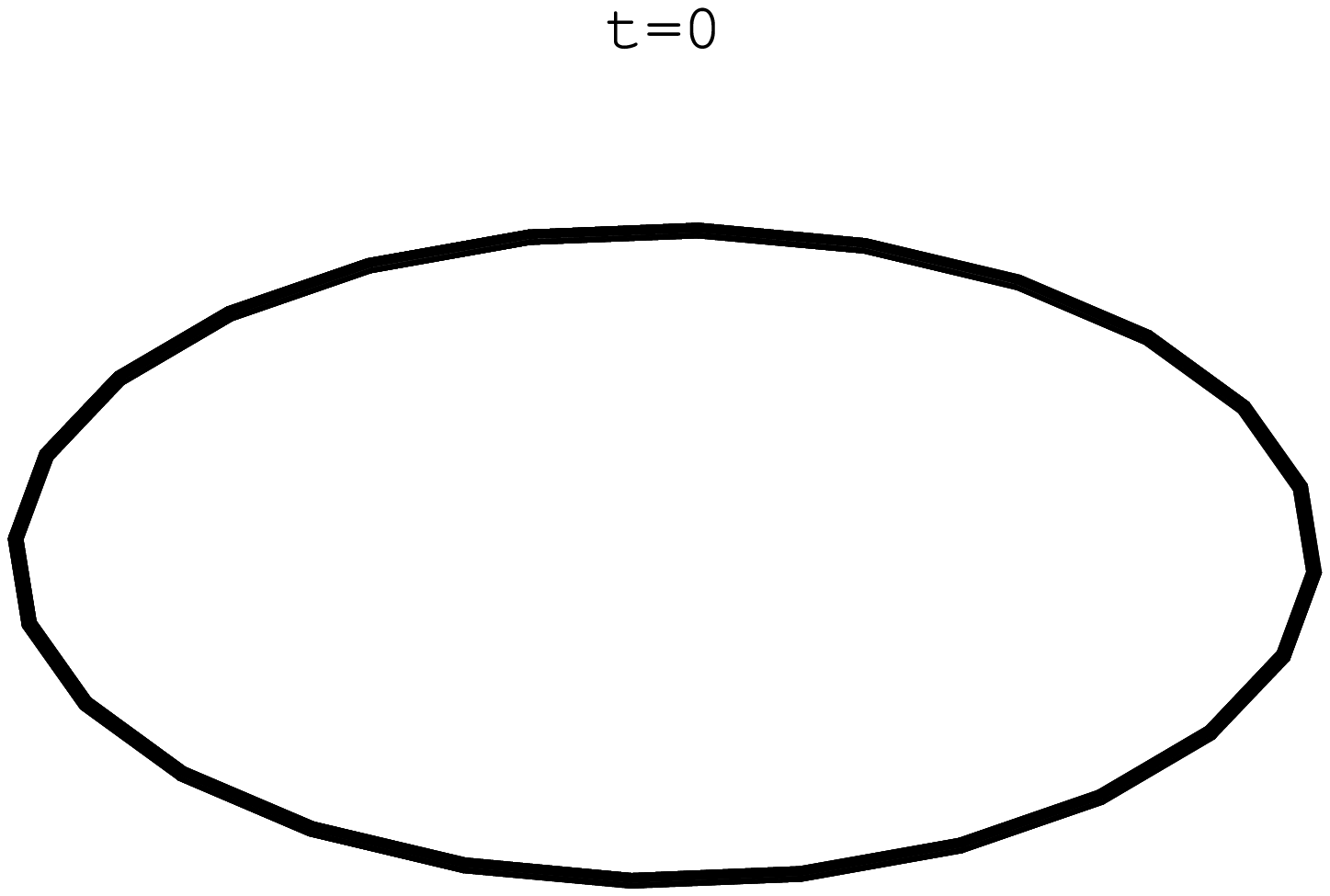, angle=-90, scale=0.4}
~~~~~~~~~~~~~~~~~~~~~~~~~
\epsfig{file=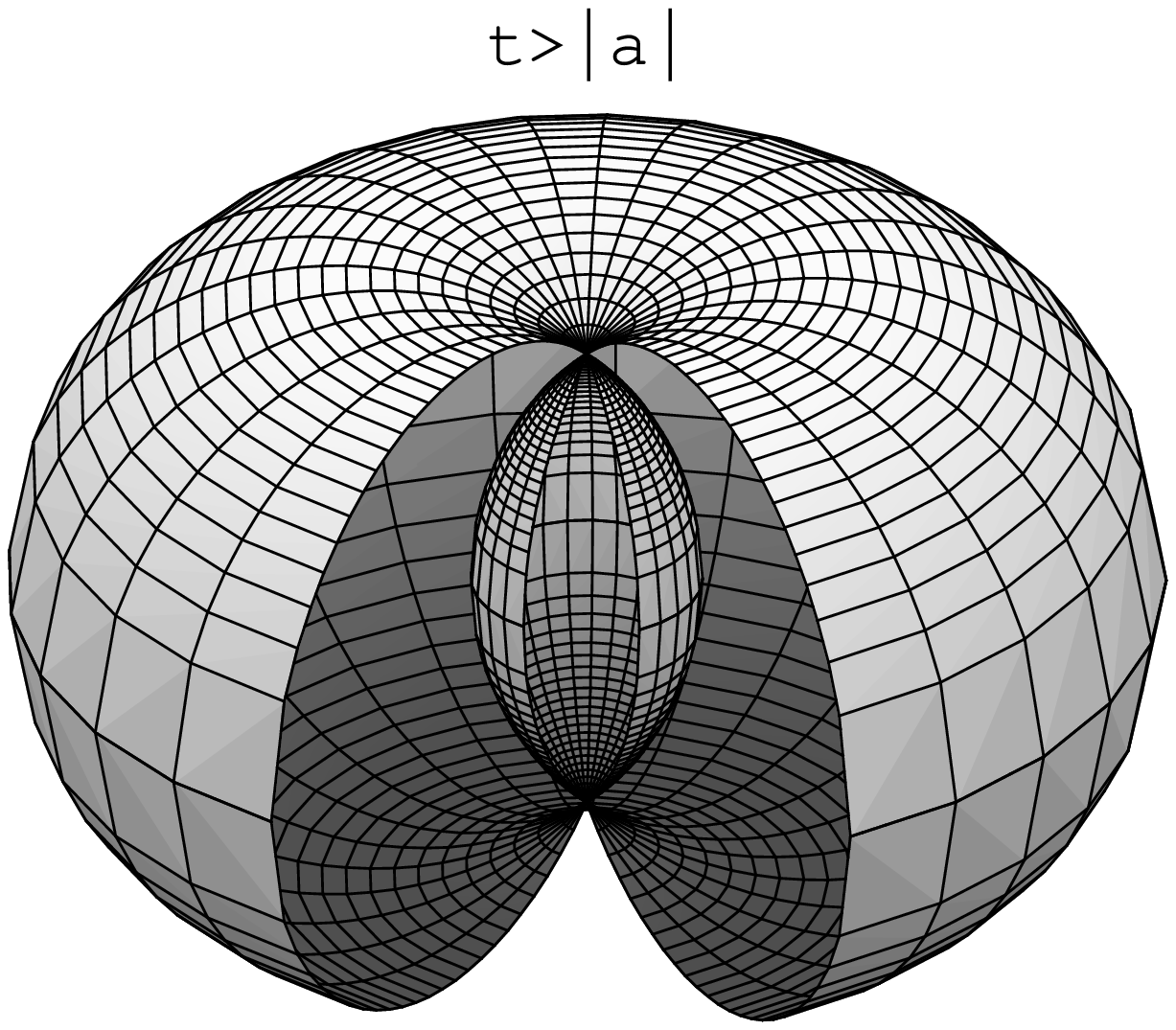, angle=-90, scale=0.4} 
\caption{Singular set of electromagnetic field (\ref{fieldborn}) for neutral 
solution (\ref{jozeph}) at initial ($t=0$) and final ($t>\vert a \vert$) stages}   
\label{pic1}
\end{figure}
\end{center}


We mention also a particle-like solution for which the singularity has a plane 
{\it 8-figure} shape at $t=0$, as well as a wave-type solution  
with a {\it helix-like} singularity~\cite{kt}. The latter stands for the 
analogue of electromagnetic waves in electrodynamics based on the GSE.

A beautiful representation of the solutions of 
SFC-equations has been proposed by E.T.Newman~\cite{newman} and developed 
then in the works A.Burinskii, R.P.Kerr and Z.Perj\'es~\cite{burin2}. 
They regarded a point-like  
source moving along some curve in a {\it complexified} Minkowski space-time 
$\mathbb{C}M$. Then a trace of its complex light cone on the {\it real} 
space-time $M$ is just a null congruence which {\it turns out to be shear-free}. 
Kerr congruence is the simplest example of this representation (a point-like 
source resting at some place of the orthogonal to $M$ {\it imaginary} subspace 
of $\mathbb{C}M$). The examples above-presented demonstrate, however, that 
the structure of singular set for such "complexified Lienard-Wiechert field" 
can be quite nontrivial. 

To conclude, a number of exact solutions of the source-free Maxwell equations 
may be obtained in a purely algebraic way, some of them being unknown before. 
They are defined exept at the points where the electromagnetic field turn to 
infinity. These points constitute a set which may be 0-, 1- or 2-dimensional 
and for a particle-like solution is bounded in 3-space. Generally, it's 
impossible to cover such set by some $\delta$-like source (because of the 
multivaluedness of the Kerr-type solutions). Nonetheless, the quantum 
numbers, 3-shape and time evolution of the singularities are well 
defined and nontrivial owing to the {\it "hidden nonlinearity"} of 
Maxwell equations in this theory, i.e. to their origination from the 
primary nonlinear GSE. The latter ensures some "selection rules" to exist 
for the solutions of Maxwell equation, in particular the restrictions 
for the allowed values of electric charge and the violation of the superposition 
principle (generally, the superposed solution may satisfy linear Maxwell 
equations but not the GSE itself). A detailed discussion of the status of 
singular particle-like solutions may be found in~\cite{kt}.


\section{Conclusion}
\label{con} 
In this paper we didn't claim to present an alternative field model or an 
algebraic method to derive the solutions of classical field theory' conventional 
equations. We only attempted to study the properties of 
differentiable functions of $\mathbb{B}$-variable themselves, i.e. to construct a  
new version of noncommutative analysis. In the particular case considered, the  
generic conditions of $\mathbb{B}$-differentiability ~\cite{kass1,kass2,kass3}  
reduce to the GSE  (\ref{GSE}) which naturally involves the gauge and the 2-spinor 
structures and manifests wonderful correlations with the structures and the language 
generally accepted in classical field theory. 

In fact, we make only three fundamental assumptions in order to physically 
interpret the abstract mathematical scheme we develop: 

\noindent 1) about space-time as a (4-dimensional subset of the) vector space 
of $\mathbb{B}$-algebra, 

\noindent 2) about physical fields as differentiable functions of 
$\mathbb{B}$-variable and 

\noindent 3) about particles as bounded singularities of 
strengths (curvatures) of gauge and metric fields naturally assosiated with 
original $\mathbb{B}$-differentiable functions-fields.
 
From the physical viewpoint, the GSE may be regarded as some peculiar 
(nonlinear, non-Lagrangian, overdetermined) field model dealing with 
effectively interacting 2-spinor and electromagnetic fields, the dynamical 
equations for the latter being not postulated but derived from the GSE itself 
as its integrability conditions. 

Twistor structure arises in the theory in a natural "dynamical" way via the 
integration of the GSE and makes it possible to obtain all its solutions, as 
well as correspondent solutions of gauge equations, in a fairly simple 
algebraic way~\footnote{In R.Penrose's twistor approach we need to integrate 
a function of twistor variable in order to get a solution to the wave equations; 
contrary, in our approach even this is not necessary.}.

Explicitly, by the examination of algebraic equation (\ref{rsol}) a wide class 
of exact solutions, of linear Maxwell equations in particular, may be found, 
solutions with bounded singularities among them. Condition 
(\ref{sin}) performs the role of {\it equation of motion} for this particle-like 
objects but at the same time predetermines the charactetristics and the spatial 
shape of singularities realizing in such a way Einstein's hypothesis of 
{\it super-causality}~\cite{einst}. 
  
Due to violation of the superposition principle for the original GSE, 
the evolution of the particle-like objects simulates physical interaction, and 
the dynamical {\it perestroikas} of the structure of singular sets may be 
interpreted as the {\it transmutations} of particles. All this reveals close 
relations to the theory of singularities of differentiable mappings and to 
the catastrophe theory~\cite{argusein}. For example, in GTR the singularity 
condition (\ref{sin}) 
is recognized as the condition of {\it caustics} of corresponding light-like 
beams formed by the shear-free congruences (\ref{cong2}).

It seems also that at least some remarkable properties of the GSE may have 
consequences for field theory in general. In particular, we refer here to \\ 
1) a possible expansion of a class of gauge models in account of weak 
gauge symmetry (\ref{tvgauge}) discovered and using the connections of the form 
(\ref{conn}),(\ref{weyl});\\
2) a natural possibility to obtain the selection rules for electric charge, 
spin etc. using the overdetermined structure of field equations;\\
3) a total algebraization of the primary PDE equipped by a twistor 
structure using the analogue of Kerr theorem or its generalizations like 
(\ref{sol}); \\
4) a possibility to establish the structure and evolution of singularities 
without even explicitly obtaining the solution of field equations itself 
(via elimination of the main field function $G(x)$ from the system of 
algebraic equations (\ref{rsol}) and (\ref{sin}), a procedure proposed 
in~\cite{kw}).

There are at least three ways to consider the material presented in the
paper and the role of GSE in particular: as a beatiful mathematical toy, 
as a powerful method to generate the solutions of conventional field equations,  
and as a fundamental dynamical system (or an example of such) primary 
to conventional Lagrangian systems. The utilization of classical 
dynamics based on the overdetermined structures like GSE requires quite new 
methods of quantization. Alternatively, one can claim to explain the quantum 
properties on the whole via, say, the stochastic behaviour of an ensemble of 
particle-like (singular) solutions or by other means of a consistently 
classical consideration. 

In any case, to find the correct approach to quantization and to physical 
interpretation in general one needs to study accurately the properties of 
the classical solutions themselves~\footnote{To clarify the correspondence 
with quantum theory the particle-like {\it multisingular} solutions are 
especially interesting.}: 
their complete classification, dynamics and bifurcations. All these  
problems are evidently related to the general theory of singularities of 
differentiable maps~\cite{argusein, arnold}. Nevertheless, the already 
discovered properties of $\mathbb{B}$-differentiable functions-fields and of 
numerous geometrophysical structures they give rise to, look like somewhat 
striking and bring one back to Pithagorean philosophy about the {\it numerical 
origin} of fundamental physical laws.

\section{Appendix}
The below reviewed approach to differentiability in the quaternion-like algebras 
have been motivated by the old works of G.Sheffers~\cite{sheff} (see also~
\cite{vishn}) on 
the analysis over an arbitrary {\it commutative} associative algebra $\mathbb{A}$, 
and is a direct generalization of Sheffer's approach to the noncommutative case. 
Let $F(Z)$  be an $\mathbb{A}$-valued function $F: \mathbb{A}\mapsto \mathbb{A}$ 
of an $\mathbb{A}$-variable $Z \in \mathbb{A}$.  
Sheffers defines the condition of its {\it differentiability in $\mathbb{A}$} 
using proportionality of the linear parts of increments (differentials) $dZ,dF$ of 
independent variable and its function respectively as
\begin{equation}\label{sheff}
dF = H(Z)*dZ ,
\end{equation}
where $H\in \mathbb{A}$ and $(*)$ denotes multuplication in $\mathbb{B}$. 
For division algebras (\ref{sheff}) is equivalent to the condition of existence 
and path-independence of the derivative $H(Z)=dF*dZ^{-1} \equiv F^\prime (Z)$, 
and in a particular case of complex algebra $\mathbb{C}$ leads to Cauchy-Riemann 
equations. Generally, however, (\ref{sheff}) may be applied to the algebras with 
zero divisors, in particular to that of the double and dual numbers. Linear 
differential equations relating the components of $F(Z)$ follow from (\ref{sheff}) 
via the elimination of $H(Z)$ and fully resemble the CR-equations for the 
functions of complex variable. In many aspects the analysis constructed by 
Sheffers is quite similar to the complex one, so that a wide class of 
$\mathbb{A}$-differentiable functions subject to (\ref{sheff}) 
can be found, including all the polinoms in particular.

In the r.h.s of (\ref{sheff}) an invariant $\mathbb{A}$-valued differential 
1-form of the most general type is present, which can be constructed via the 
algebraic operations in $\mathbb{A}$ only. A natural generalization of (\ref
{sheff}) to the case when $\mathbb{A}$ is noncommutative (yet associative) 
seems to be the following condition (see~\cite{kass1,kass2} and the references 
therein):
\begin{equation}\label{ncd}
dF = L(Z)*dZ*R(Z)
\end{equation}
for the function $F(Z)$ to be differentiable in $\mathbb{A}$. Here $L,R
\in \mathbb{A}$ are the so called left and right {\it semi-derivatives} of $F(Z)$ 
respectively. For a given function $F$ they are determined (if exist) not 
uniquely, but at least up to a transformation $L\rightarrow \alpha L,~R\rightarrow 
\alpha^{-1} R$,  where the function $\alpha(Z)$ takes the values in  
the {\it centre} (the commutative subalgebra) of $\mathbb{A}$. 

For commutative algebras (\ref{ncd}) reduces back to (\ref{sheff}) with 
$H(Z)\equiv L(Z)*R(Z)$ . On the other hand, if in noncommutative case we take, 
say, $R=e$  (we assume the unit element $e$ to exist in $\mathbb{A}$), 
we come to the condition (\ref{sheff}) with $H(Z)\equiv L(Z)$. 
Nonetheless, at least for the algebras of quaternion type this condition is 
known to be too restrictive, being satisfied only by the linear function 
$F=A*Z+ B$, where $A,B$ are constant elements of algebra (see e.g. 
~\cite{sud, dev}).

As to the general $\mathbb{A}$-differentiability condition (\ref{ncd}), it 
defines a wider class of functions. Particularly, for Hamilton quaternions  
$\mathbb{H}$ condition (\ref{ncd}) turns out to be an algebraic analogue of  
for the mapping $Z\mapsto F(Z)$ {\it to be conformal} in $E^4$ ~\cite{kass5,kass4,
kass3}. In this regard equation (\ref{ncd}) can be viewed at as a natural 
generalization of complex holomorphy. 
However, in $E^4$  conformal mappings constitute only a finite 
15-parameter group, contrary to the infinite-dimensional complex case. Thus, 
for division algebra $\mathbb{H}$ the class of $\mathbb{H}$-differentiable 
functions, as defined by (\ref{ncd}), is again too narrow to be used, say, in 
field dynamics.

The situation changes radically when we come to consider noncommutative 
algebras with zero divisors, in particular the algebra of biquaternions 
$\mathbb{B}$ (quaternions over $\mathbb{C}$). For simplicity let us below 
restrict ourselves by consideration of the full $N\times N$ matrix algebras 
over $\mathbb{R}$ or $\mathbb{C}$ (for $N=2$ we have just an isomorphism 
$Mat(2,\mathbb{C})\cong \mathbb{B}$). Then for a {\it determinant} of the 
matrix of differentials $dF$ in the l.h.s. of (\ref{ncd}) we obtain
\begin{equation}\label{norm}
\det\Vert dF \Vert = \det\Vert L(Z)*R(Z)\Vert \det\Vert dZ\Vert \equiv 
\lambda(Z) \det\Vert dZ \Vert.
\end{equation}
In the case both $L,R$ are invertible, so that $\lambda(Z)\neq 0$, relation 
(\ref{norm}) defines a conformal mapping with a scale factor $\lambda(Z)$ 
and a (positively indefinite or complex) infinitesimal metric {\it interval} 
represented by corresponding determinants in (\ref{norm}). Particularly, for  
$\mathbb{B}$ we deal with conformal mappings in complexified Minkowski space 
$\mathbb{C}M$. 

In a remarkable way, however, for $\det L =0$ or similarly $\det R = 0$, 
we have $\lambda(Z)=0$ and relation (\ref{norm}) defines a reduction of the full 
vector space of $\mathbb{A}$ to the subspace of null elements (to the complex 
"light cone" for $\mathbb{B}$). Such mappings may be called {\it degenerate conformal 
mappings}; they constitute a wide and important class: in the context of the 
presented theory just these mappings (differentiable $\mathbb{A}$-functions) are 
identified with the physical fields.

In the $N\times N$ matrix notation (\ref{ncd}) takes the form ($A,B,...=1,...N$)
\begin{equation}\label{matnot}
\nabla_{AB}F_{CD} = L_{CA}R_{BD}
\end{equation}
where $\nabla_{AB}$ stands for a derivative operator with respect to the 
coordinate $Z^{AB}$. For some indices $C,D$ being fixed we denote $F_{CD}\equiv 
\Sigma, ~L_{CA}\equiv \phi_A, ~R_{BD}\equiv \psi_B$, and the relation (\ref{matnot}) 
becomes
\begin{equation}\label{matnot2}
\nabla_{AB}~ \Sigma = \phi_A\psi_B
\end{equation}

In view of the zero determinant of the matrix in the r.h.s., we get 
the equation 
\begin{equation}\label{eikonal1}
\det\Vert \nabla_{AB} \Sigma \Vert = 0, 
\end{equation}
which have to be satisfied for each matrix component $F_{CD}\equiv \Sigma \in 
\mathbb{R}$, $\mathbb{C}$ of an $\mathbb{A}$-differentiable function. Equation (\ref
{eikonal1}) is a {\it nonlinear} analogue of the Laplace equation in the 
complex analysis, and nonlinearity arises here as a direct consequence of the 
account of noncommutativity in the definition of a differentiable function 
(\ref{ncd}). For the case of biquaternions $\mathbb{B}$ (\ref{eikonal1})
represents a (complexified) 4-{\it eikonal} equation 
\begin{equation}\label{ceicon}
(\nabla_{00}\Sigma)(\nabla_{11} \Sigma) - (\nabla_{01} \Sigma) 
( \nabla_{10} \Sigma) = 0 
\end{equation} 
which in Cartesian complex coordinates $z^0,z^3=z^{00}\pm z^{11},~
z^1,z^2=z^{01}\pm iz^{10}$ takes a familiar form
\begin{equation}\label{fameikon}
(\frac{\partial \Sigma}{\partial z^0})^2 - (\frac{\partial \Sigma}
{\partial z^1})^2 - (\frac{\partial \Sigma}{\partial z^2})^2 - (\frac{\partial 
\Sigma}{\partial z^3})^2 = 0 . 
\end{equation}

In the paper we restrict our consideration by a particular, yet a basic subclass 
of $\mathbb{A}$-differentiable functions for which one of semi-derivatives, 
say $R(Z)$, is proportional to the function $F(Z)$ itself. Redefining then  
$L(Z)\rightarrow \Phi(Z)$ we get instead of (\ref{ncd})
\begin{equation}\label{covconst}
dF = \Phi(Z)*dZ*F(Z) .
\end{equation}

Let now $\{\xi^{(C)}\},~C=1,...N$ be $N$ columns of the matrix $F(Z)$; then 
we can present (\ref{covconst}) in a form of a system of $N$ matrix equations 
\begin{equation}\label{covspin}
d\xi^{(C)} = \Phi dZ \xi^{(C)}
\end{equation}
(here and below we omit the symbol of matrix multiplication),
which all have to be satisfied with the same (left semi-derivative) 
matrix $\Phi(Z)$. The quantities $\xi (Z)$ are evidently $SL(N,\mathbb{C})
$-spinors with respect to the symmetry transformations of (\ref{covspin}) 
(for $N=2$ we have demonstrated this in section 2).

Different spinors $\{\xi^{(C)}\}$ may be either functionally dependent or not
\footnote{In the latter case we return back to the conformal maps in 
the $N^2$ vector space of $\mathbb{A}$, with the norm represented by the 
determinant.}: in any case an arbitrary solution of (\ref{covconst}) may be 
constructed from (and decomposed into) a set of $N$ solutions $\{\Phi(Z), 
\xi^{(C)}(Z)\}$ of the system
\begin{equation}\label{GSE23}
d\xi = \Phi dz\xi .
\end{equation}
Conversely, from a solution to (\ref{GSE23}) we easily obtain at least 
one class of the solutions to the original system (\ref{covconst}) by 
setting $N$ spinors $\xi^{(C)}(Z)$ to be globally proportional to each 
other (or zero, exept for one of them). Eliminating the semi-derivative 
matrix $\Phi$ from the overdetermined system (\ref{GSE23}) we come to a  
nonlinear system for the components of $\xi$ (i.e. of 
$\mathbb{B}$-differentiable function) only which resembles in a certain 
sense the Cauchy-Riemann conditions in the complex calculus (for $N=2$ this 
is just the system (\ref{rbccv}) of the paper). 

Thus, we have shown that (in a particular case $R(Z)=F(Z)$ ) the $\mathbb{A}$-
differentiable functions, as defined by (\ref{ncd}), can all be found from the 
condition (\ref{GSE23}). For the algebra of biquaternions $\mathbb{B}$ this 
equation becomes equivalent to the GSE (\ref{GSE}) studied in the paper if 
we only assume the coordinates to be {\it real-valued}, so that the matrix 
$Z \rightarrow X, ~X=X^+$ is considered to be hermitian. 

This is the only 
{\it ad hoc} assumption which is motivated by physical considerations and 
which does not follow from the algebraic structure or the differentiability 
conditions (\ref{ncd}) themselves (precisely, we have to deal with the 
full structure of 4-dimensional complex space). 
Under the assumption made, the coordinate space reduces to the Minkowski 
one and the whole theory, including the fundamental 4-eikonal equation 
(\ref{fameikon}), becomes Lorentz invariant.

Some details and generalizations of the approach afore-presented can be 
found in the monograph ~\cite{kass1} and, partly in English, in ~\cite{kass5,
kass2,kass3}.

\end{document}